# Emergence of equilibrium thermodynamic properties in quantum pure states I. Theory.

**Barbara Fresch[a], Giorgio J. Moro[b]**

*Department of Chemical Science, University of Padova, Via Marzolo 1, 35131 Padova – Italy*

**Abstract**

Investigation on foundational aspects of quantum statistical mechanics are recently entered in a renaissance period due to novel intuitions from quantum information theory and an increasing attention to the dynamical aspects of single quantum systems. In the present contribution a simple but effective theoretical framework is introduced to clarify the connections between a purely mechanical description and the thermodynamic characterization of the equilibrium state of an isolated quantum system. A salient feature of our approach is the very transparent distinction between the statistical aspects and the dynamical aspects in the description of isolated quantum systems. Like in the classical statistical mechanics, the equilibrium distribution of any property is identified on the basis of the time evolution of the considered system. As a consequence equilibrium properties of quantum system appear to depend on the details of the initial state due to the abundance of constants of the motion in the Schrödinger dynamics. On the other hand the study of the probability distributions of some functions, such as the entropy or the equilibrium state of a subsystem, in statistical ensembles of pure states reveals the crucial role of typicality as the bridge between macroscopic thermodynamics and microscopic quantum dynamics. We shall consider two particular ensembles: the Random Pure State Ensemble and the Fixed Expectation Energy Ensemble. The relation between the introduced ensembles, the properties of a given isolated system and the standard quantum statistical description is discussed throughout the presentation. Finally we point out the conditions which should be satisfied by an ensemble in order to get meaningful thermodynamical characterization of an isolated quantum system.

## I. INTRODUCTION

Recently, the possibilities of investigating single molecule, or single spin observables, as well as the necessity of a better understanding of the mechanisms underlying quantum dynamics in order to obtain nanoscale devices and nanostructured materials suitable for quantum computing tasks, have revived the interest in foundational aspects of quantum statistical mechanics. The impressive development of quantum information theory [1] has changed the perceptions of quantum entanglement: for a long time it has been considered a somewhat paradoxical property of the matter at the atomic scale, but now it is regarded as an essential and ubiquitous phenomenon whose consequences are affecting the very macroscopic world that we experience [2]. Still its role in the foundations of statistical physics has been only recently recognized [3]. On the other hand, the *decoherence program* initiated by Zurek [4] has provided a better understanding of the measurement process as well as the transition from quantum to classical world. By avoiding the postulated "collapse" of the wavefunction, the studies on decoherence has moved the attention back again to the unitary evolution of the total wavefunction by demonstrating how the entangling interactions between a quantum system and its environment can destroy coherence amongst the states of the quantum system. While the question whether decoherence provides or at least suggests a solution to the measurement problem is still object of debate and active research [5], it is sure that decoherence brings out the importance of considering the evolution of a quantum system together with its surrounding in order to clarify key aspects of the quantum theory.

[a] barbara.fresch@unipd.it

[b] giorgio.moro@unipd.it

It is then important to base the analysis on the behaviour of the isolated system including the system of interest and its environment as well [6,7,8]. Only for the isolated system, which is characterized by a well defined wavefunction in opposition to open quantum systems, one can study the dynamics by solving the Schrödinger equation [9] once the Hamiltonian is given. Recently this standpoint has led to a paradigmatic change on the foundations of quantum statistical mechanics: from the description of the averages on an ideal collection of many independent copies of the given quantum system through the statistical density matrix [10,11], to the statistical characterization of the wavefunction of an isolated quantum system [3,12,13,14]. In the latter framework, a key ingredient is the probability distribution on the wavefunction of the isolated system (or better, the probability density on the stochastic variables parameterizing such a wavefunction), which has to be postulated on the basis of some general hypothesis about the symmetry and the constraints of the wavefunction. The idea of assigning probability distribution to wavefunctions has been used in the past to calculate on a statistical basis molecular properties such as transition probabilities [15,16] and to characterize vibrationally excited states of polyatomic molecules [17,18]. Once such a probability density is provided one can evaluate the statistical properties of any observable like the expectation value of relevant operators or any function of the quantum state. In particular, differently from conventional quantum statistical mechanics which provides only the average of the observables by means of the thermal density matrices (microcanonical or canonical), this approach leads to a probability distribution on the observables. Such a feature, which might appear as a drawback for the absence of precise predictions on the observables, has not significant effects as long as the statistics manifests typicality [19]. In its widest meaning the term typicality indicates that by selecting the allowed states on the basis of some conventional rules, one obtains a very narrow distribution of some relevant features, which become, actually, typical among these states. As shown by Goldstein et all. [12] and by Popescu et all. [3], in the asymptotic limit for increasing size of the Hilbert space associated to the system wavefunction, almost all the possible pure states lead to the same reduced density matrix of a small subsystem. In such a condition the distribution on the states of the subsystem has no detectable effects due to its typicality. Subsequently, Reimann [13] has shown that the concept of typicality can be applied to a wider class of observables, for instance collective properties of the isolated quantum system, which are described by operators with suitable boundaries on their spectrum. In conclusion, the statistical description of the pure states of an isolated system is capable of providing definite predictions on the observables as long as these become typical in suitable conditions.

An important point to be stressed is that typicality analysis requires the preliminary choice of a well defined statistical distribution for the pure state of the isolated system [3,12,13]. The most general analysis by Popescu et al. assumes an homogeneous distribution for the wavefuctions belonging to a subspace of the entire Hilbert space defined by an arbitrary constraint. Since in the framework of statistical thermodynamics the relevant constraint is the total energy, Goldstein et al. [12] consider wavefunctions belonging to the subspace defined by the Hamiltonian principal directions with eigenvalues in a small shell between $E - \Delta E$ and $E$. An alternative ensemble is generated by substituting this constraint on the allowed Hamiltonian eigenvalues, with the constraint for the expectation value of the Hamiltonian $E = \langle \psi | H | \psi \rangle$ and the wavefunction represented in the full Hilbert space [20,21,22]. It is worth to note that in this framework, and in the analysis we present in the following, the concept of ensemble is intended in its formal statistical meaning [50], that is the sample space for the set of stochastic variables together with the probability density for them. Therefore it should not be strictly identified with a collection of copies of the isolated system.

Both the previously mentioned ensembles are obtained by considering the wavefunction as a geometrical entity without taking explicitly into account its unitary time evolution. In the present communication we propose a different scheme for the statistical description of the wavefunction of an isolated quantum system. Like in classical statistical mechanics [63], one can follow the time evolution of the isolated system in order to reconstruct the equilibrium distribution of the possible quantum states. We shall derive explicitly under suitable conditions such a distribution function, in the following called as Pure State Distribution (PSD), since it is determined by the dynamics of a given pure state. The connection with the statistical description of classical system results even more stringent by tacking into account that the Schrödinger evolution is formally equivalent to the Hamilton equation of motion for an ensemble of harmonic oscillators [23]. Then, in strict analogy with the standard procedure of classical statistical mechanics, the equilibrium value of a given observable is identified with its asymptotic time average, which can be evaluated as the average on the PSD. It should be stressed that the Pure State Distribution allows the calculation not only of the equilibrium average of a given property, but also of its higher moments, for instance its variance which quantifies the time fluctuations of the observable with respect to its average [56].

It is easily shown that the sample space for the PSD is given by the phases of the components of the wavefunction along the Hamiltonian principal directions, with the corresponding squared norms identified by us with the populations, playing the role of constants of motion. This implies that the equilibrium average, or more generally the entire distribution, of any function of the quantum state, depends on the population set. Furthermore the Schrödinger dynamics of the wavefunction does not provide any information on such a dependence on the populations.

Evidently, quantum mechanics does not provide definite rules for a univocal choice of the populations. Therefore we can describe them only within a statistical framework, in the same spirit of the previous studies on typicality by Goldstein et al [12] and by Riemann [13]. Like in these references, an ensemble can be proposed on the basis of the symmetry of the wavefunction and on the constraint for the energy of the isolated system. However, while the ensembles used in these previous works can be reformulated as a statistical distribution on the sample space of both phases and populations, our ensembles refer to the sample space of the populations only. This is a direct consequence of employing the Schrödinger dynamics to determine the distribution on the phases, without the need of any a priori assumption about the randomness of these variables. More specifically we shall consider two different ensembles: the Random Pure State Ensemble (RPSE) which might be seen as a generalization of the ensemble introduced in ref. [12] without the limitation of the small energy window for the participating eigenstates, and the Fixed Expectation Energy Ensemble (FEEE), where the constraint on the energy $E$ of the isolated system is introduced through the expectation value of the Hamiltonian, $E = \langle \psi | H | \psi \rangle$.

The same fact that populations can be characterized only statistically implies a statistical distribution on the quantum equilibrium properties as derived from PSD. Thus, the issue of attributing well defined values to equilibrium properties, like in the thermodynamical description of macroscopic systems, naturally arises. The solution of this problem is found by means of the concept of typicality in the same spirit of refs. [3,12-14], even if we consider ensembles of a different nature with the sample space including only the populations of the pure state. Our analysis is formulated by considering first systems of finite size, such that the typicality of a property can be quantified by evaluating the variance of its distribution within a particular ensemble. Then, the infinite size limit of the system is considered, i.e. the thermodynamic limit, in order to verify whether a specific property is characterized by a strong form of typicality in correspondence of a negligible effect of the variance.

The previous procedure requires to consider the following preliminary issue: how to choose the correct statistical ensemble for the populations? It should be evident that a similar question arises also for the statistical ensemble of pure states as considered in refs. [3,12-14]. Indeed the typical reduced density matrix depends on the global constraints on the Hilbert space, as clearly pointed out in the analysis of Popescu et al.

As long as there are not logically necessary, a priori criteria for an univocal choice of the ensemble, one has to impose some a posteriori conditions on the agreement of calculated typical values of observables with the known behaviour of material systems. In this framework we shall impose the condition that, in the asymptotic limit of infinitely large systems, the equilibrium properties of the isolated system should have the formal structure of state functions like in equilibrium thermodynamics. Furthermore, we also require that the typical reduced density matrix describing a subsystem has the standard canonical form at the temperature determined by the entropy equation of state of the overall isolated system. We shall show that these conditions effectively discriminate the two ensembles explicitly considered here (RPSE and FEEE)

Our contribution is articulated in two papers. In this first part of the work we shall present the theoretical framework which permits to establish the connection between macroscopic thermodynamics and the detailed characterization of a quantum pure state. In particular, in Section II we clearly define the concept of equilibrium and derive the Pure State Distribution on the basis of the Schrödinger evolution. In Section III, two different statistical ensembles of pure states are introduced together with the corresponding probability density functions. The relation between these ensembles, the properties of the time evolving single system and the standard statistical description is discussed throughout the presentation. Finally, in Section IV we point out the conditions under which a meaningful thermodynamical characterization of an isolated quantum system in a pure state can be given.

The second paper, in the following denoted as part II, is devoted to the application of the presented scheme to a model system. This second step, besides being relevant in order to verify the applicability of the developed theoretical framework, is necessary in order to establish the different characteristics of the two particular statistical ensembles we have considered.

## II. EQUILIBRIUM DISTRIBUTION FROM THE DYNAMICS OF QUANTUM PURE STATES

We shall introduce the equilibrium properties and the probability distribution for a quantum system on the basis of its dynamics. In the quantum domain the time evolution is ruled by the Schrödinger equation which determines a unitary dynamics for the wavefunction. Like the classical Hamiltonian equations, the Schrödinger equation refers to an isolated system. However, the condition of isolation is by far more stringent in quantum mechanics then in the classical case: indeed, even in the absence of any energetic interactions, the system can always be entangled with another quantum system with which it has interacted in the past [24,25,26]. In this case it is not possible in general to assign two separate wavefunctions to the two systems. In the following we intend as isolated a quantum system which is in a pure state (i.e., describable by a wavefunction) and, therefore, without interaction and entanglement with the surrounding.

## A. Quantum evolution and equilibrium properties

Let $E_n$ and $|e_n\rangle$ denote the eigenvalues and the eigenfunctions of the system Hamiltonian $H$ which is assumed to be time independent since the system is isolated. In the energy representation the wavefunction reads

$$\psi(t) = \sum_{n=1}^{N} c_n e^{-iE_n t/\hbar} |e_n\rangle \tag{1}$$

where $c_n := \langle e_n | \psi(0) \rangle$ are the components of the initial state in the energy representation and $N$ denotes the dimension of the Hilbert space $\mathcal{H}$ which is assumed to be finite. A completely equivalent description of the quantum system is obtained by means of the corresponding pure state density operator $\rho(t)$, which is defined as the projection operator onto the one dimensional subspace determined by $\psi(t)$

$$\rho(t) = |\psi(t)\rangle\langle\psi(t)| = \sum_{n,m=1}^{N} c_n c_m^* e^{-i(E_n - E_m)t/\hbar} |e_n\rangle\langle e_m| \tag{2}$$

It determines the instantaneous expectation value $a(t)$ of any quantum mechanical operator $A$ as

$$a(t) = \langle \psi(t) | A | \psi(t) \rangle = \mathrm{Tr}\{A\rho(t)\} \tag{3}$$

In the essence, from the knowledge of the Hamiltonian and of the initial state $\rho(0)$ of the system, one predicts, in the deterministic meaning, the state $\rho(t)$ of such a system at any time and in the full detail. Correspondingly, the time dependent observable would be recovered, with a fluctuating profile due to the superposition of many oscillatory contributions brought by the matrix elements of $\rho(t)$. Like in classical statistical mechanics [63], we introduce the asymptotic time average to be identified with the equilibrium value of the property

$$\overline{a} = \lim_{T \to \infty} \frac{1}{T} \int_0^T dt\, a(t) = \mathrm{Tr}(A\overline{\rho}) \tag{4}$$

Correspondingly, the averaged density matrix $\overline{\rho}$

$$\overline{\rho} = \lim_{T \to \infty} \frac{1}{T} \int_0^T dt\, \rho(t) \tag{5}$$

should be identified as the appropriate object to compare with the notional equilibrium density matrix. Hereafter, the over-bar on a given property will denote its equilibrium value, to be evaluated as the asymptotic time average of the pure state evolution. In the absence of any degeneracy in the energy spectrum, one has that $\lim_{T\to\infty} \frac{1}{T}\int_0^T dt\, e^{-i(E_n-E_m)t/\hbar} = \delta_{n,m}$, and the time average is readily performed to obtain

$$\bar{\rho} = \sum_{n=1}^{N} |c_n|^2 |e_n\rangle\langle e_n| = \sum_{n=1}^{N} P_n |e_n\rangle\langle e_n| \tag{6}$$

where we have introduced the parameters $P_n = |c_n|^2$ for $n = 1, 2, \cdots, N$, which we shall call "populations", each of them being associated to a particular Hamiltonian eigenstate. They are the diagonal elements of the pure state density matrix

$$P_n = \mathrm{Tr}\left(|e_n\rangle\langle e_n| \rho(t)\right) = \rho_{nn} \tag{7}$$

which do not depend on time since $\left[H, |e_n\rangle\langle e_n|\right] = 0$ and, therefore, they represent constants of motion for the quantum problem.

The time evolving quantum state, eq. (1), can be parameterized in terms of a set of constant populations $P = (P_1, ..., P_N)$ and time dependent phases $\alpha(t)$ with $\alpha = (\alpha_1, \alpha_2, \cdots, \alpha_N)$. Indeed, by writing the expansion coefficients in their polar form as $c_n = \sqrt{P_n} e^{i\alpha_n(0)}$, the instantaneous wavefunction reads

$$\psi(t) = \sum_{n=1}^{N} \sqrt{P_n} e^{-i\alpha_n(t)} |e_n\rangle \tag{8}$$

where the $N$ phases are linear functions of time

$$\alpha_n(t) = -\alpha_n(0) + E_n t/\hbar \tag{9}$$

Of course the populations and the phases parameterize also the density operator

$$\rho(t) = \sum_{n,m} \sqrt{P_n P_m} e^{-i\alpha_n(t) + i\alpha_m(t)} |e_n\rangle\langle e_m| \tag{10}$$

Due to the normalization of the wavefunction, $\langle\psi|\psi\rangle = 1$, the set of populations fulfills the normalization requirement

$$\sum_{n=1}^{N} P_n = 1 \tag{11}$$

Moreover they determine the expectation energy of the system

$$E = \langle\psi|H|\psi\rangle = \sum_{n=1}^{N} P_n E_n \tag{12}$$

as well as the entropy given by

$$S = -k_B \sum_{n=1}^{N} P_n \log P_n \tag{13}$$

where $k_B$ is the Boltzmann constant. The entropic function defined in the above equation depends only on populations and is characteristic of a given quantum pure state. It is not the von Neumann entropy [27], $-\text{Tr}\{\rho(t)\ln\rho(t)\}$, which for a pure state is always zero. It actually corresponds to the entropy as defined by Shannon [28], which is usually interpreted as a measure of the lack of information about the outcome of a measurement of the energy. In what follows we do not consider any measurement process, so the function (13) is rather interpreted as a measure of the degree of disorder of a quantum pure state with respect to its decomposition onto the Hamiltonian eigenstates. In particular vanishing entropy would be recovered only for a stationary eigenenergy state, $\psi(t) \propto e_k$ for a given $k$.

Such a framework allows also the investigation of the behaviour of a subsystem $S$ of the entire system $S+E$, where $E$ denotes generically the environment. Since our analysis is primarily based on the time evolution determined by the Schrödinger equation, we will always consider the total (isolated) system $S+E$ to be in a pure state described by a wavefunction $\psi(t)$. Accordingly, the state of the subsystem has to be specified by means of its reduced density operator which is defined as the partial trace over the environment of the pure state density operator eq. (2)

$$\mu(t) = \text{Tr}_E\{\rho(t)\} \tag{14}$$

Like for the generic observable eq. (3), a fluctuating behaviour is expected for the elements of the reduced density matrix, with their equilibrium value derived through the asymptotic time average

$$\bar{\mu} = \lim_{T\to\infty} \frac{1}{T}\int_0^T dt\, \text{Tr}_E(\rho(t)) = \text{Tr}_E(\bar{\rho}) \tag{15}$$

Notice that the equilibrium properties of a generic observable include not only its average

$$\bar{a} = \text{Tr}(A\bar{\rho}) = \sum_n A_{nn} P_n \tag{16}$$

with $A_{mn} = \langle e_m | A | e_n \rangle$, but also its fluctuations. The fluctuation amplitude is given by the average $\overline{\Delta a^2}$ of the squared deviation $\Delta a(t) = a(t) - \bar{a}$, which is explicitly given by

$$\overline{\Delta a^2} = \sum_{n'\neq n} A_{nn'} A_{nn'}^* P_n P_{n'} \tag{17}$$

as long as the transition frequencies of the Hamiltonian are non degenerate. Equilibrium average properties are thus given as explicit functions of the of the populations. In particular we shall focus on the average reduced density matrix, eq. (15), on the expectation energy, eq. (12) and the entropy, eq. (13), which are also determined by the population set.

## B. Motion in the Hilbert space: Pure State Distribution

Given the parameterization eq. (8) of the wavefunction, we can consider the phases $\alpha(t)$ as the statistical variables which determine the instantaneous state of the system, for a given set of populations $P$. Then, the distribution of the phases during the time evolution of the system allows us to introduce the conditional probability density $p(P|\alpha)$ on the phases for given populations, with normalization

$$\int d\alpha\, p(P|\alpha) = 1 \tag{18}$$

where $\int d\alpha := \int_0^{2\pi} d\alpha_1 \int_0^{2\pi} d\alpha_2 \cdots \int_0^{2\pi} d\alpha_N$, having chosen the standard definition domain for each angle: $0 \leq \alpha_j < 2\pi$. We shall call such a probability density as the Pure State Distribution (PSD). It can be derived by imposing the condition that, for any function of the phases $f_P(\alpha)$, in general parametrically dependent on the populations, the phase average according to $p(P|\alpha)$ and the asymptotic time average along a trajectory should be equivalent:

$$\overline{f_P(\alpha(t))} = \lim_{T\to\infty} \frac{1}{T} \int_0^T dt\, f_P(\alpha(t)) = \int d\alpha f_P(\alpha) p(P|\alpha) \tag{19}$$

From simple considerations about the phase variables eq. (9) one derives that, under the condition of rational independence [29] of the eigen-energies (also called non-resonance condition), the Pure State Distribution is independent of the populations and it can be represented as an homogeneous function of the phases [30]:

$$p(P|\alpha) = \frac{1}{(2\pi)^N} \tag{20}$$

The proof is given in Appendix A by tacking into account that the eigen-energies are rationally independent if the equation

$$\sum_{j=1}^{N} n_j E_j = 0 \tag{21}$$

with $n_j$ being integers, has only the trivial solution $\left(n_j = 0,\ \forall j\right)$.

We emphasize that the non resonance condition of the eigen-energies [31] is a natural and weak restriction for real systems. Indeed, the presence of an arbitrarily small random perturbation in the Hamiltonian [32,33,34] actually leads to a rationally independence energy spectrum.

Then, if models (like the so called ideal systems) lacking of the property of eigen-energy independence are examined, as it will be done by us in part II, they should be intended as an approximation of real systems neglecting a variety of weak interactions. Such an approximation allows a simple characterization of the energy spectrum of the system but it does not question the existence of the corresponding Pure State Distribution.

## C. Connections with the Ergodic Problem and Standard Statistical Density Matrices

The Pure State Distribution reflects the characteristics of the unitary quantum evolution: it is well known that for a given set of populations, the associated quantum mechanical motion is equivalent to a uniform translation on a torus of appropriate dimensionality [35]. The condition of rational independence of eigenenergies ensures that such a torus is covered ergodically in the meaning given by the classical Birkoff's theorem which is reported in the appendix B for the sake of completeness, together with the definition of the key concept of metrically indecomposable subspace. The identification of the N-dimensional torus as the ergodic subspaces of the Hilbert space also follows from the Birkoff's theorem formulated for the phase space $\Gamma$ of a classical system once established the formal analogy between Schrödinger equation and Hamilton classical equation of motion. Let us consider the expansion of the wavefunction on the energy basis eq. (1) and let us associate the real and imaginary parts of the $n$-th coefficient $c_n(t) = \langle e_n | \psi(t) \rangle$ to the $n$-th coordinate and momentum, respectively, $q_n = \operatorname{Im} c_n$ and $p_n = \operatorname{Re} c_n$, for $n = 1, 2, \cdots N$. Then, the Schrödinger evolution of the wavefunction is equivalent to the following classical (Hamiltonian) equation of motion [36]

$$\frac{dq_n}{dt} = \frac{\partial E(q,p)/2\hbar}{\partial p_n} \qquad \frac{dp_n}{dt} = -\frac{\partial E(q,p)/2\hbar}{\partial q_n} \qquad (22)$$

where

$$E(q,p) = \sum_{k=1}^{N} E_k (p_k^2 + q_k^2) \qquad (23)$$

Since these are the equations of motion in the phase space of $N$ classical harmonic oscillators, any wavefunction of the form (1) can be represented as a point in the $2N$ dimensional classical phase space. Then, the change of coordinates $(\operatorname{Im} c, \operatorname{Re} c) \to (P, \alpha)$ in such a space allows the identification of the conserved quantity, i.e. the set of populations.

In the classical case one can conjecture that by adding suitable interactions to the Hamiltonian of harmonic oscillators, the entire constant energy surface will become metrically indecomposable [37,38], which means that there must be no dynamical variables, other than functions of the Hamiltonian, that are constants of the motion (as indicated in figure 1A). If such condition can be proven, then the Birkhoff's theorem assures the equivalence between the infinite time average and the average over the constant energy surface (i.e. the microcanonical average) and the system is said to be *ergodic* [39]. More often, a rigorous proof of such a property is not possible and one tacitly *assume* the validity of an "ergodic hypothesis" [40] in order to justify the use of the microcanonical distribution. On the contrary, there is no room for such a procedure in the quantum mechanical framework. Indeed the identification of the ergodic subspaces cannot lead in any way to the justification of the microcanonical statistics since different sets of population always identify different *metrically indecomposable* regions of the entire Hilbert space (as represented in part B of figure 1).

From this perspective the statistical description of the equilibrium of the isolated quantum system is recovered from the particular Pure State Distribution once the populations are selected. The conventional quantum microcanonical statistics [41,42,43] is then associated to the average density matrix for a particular choice for the populations, when they are all equal within an egein-energy domain specified by its boundaries $E_{\min}$ and $E_{\max}$, that is

$$P_n = \begin{cases} \dfrac{1}{N} & \text{if} \quad E_{\min} \leq E_n \leq E_{\max} \\ 0 & \text{otherwise} \end{cases} \qquad (24)$$

where $N$ is the number of populated eigen-energy states. However, it should be emphasised that the particular PSD obtained by using the model for the populations given in eq. (24) is conceptually different from the conventional quantum microcanonical statistics. Indeed the latter concerns only with the average density matrix, while the PSD is intended to describe the full distribution on the instantaneous values of the density matrix. Recently a generalized microcanonical ensemble has been proposed by Brody et al. [44,45] in which the microcanonical energy is identified with the expectation value of the Hamiltonian. This framework defines an *ensemble of pure states* since many different sets of populations can give the same expectation energy. On the other hand, it is not clear how such an ensemble description can be related with the statistical characterization of a single quantum state. To circumvent this point Naudts et al. [46] proposed the application of the maximum entropy principle. In short one can maximize the entropy as defined in eq. (13), under the constraints of normalization of the populations and constant expectation energy, eq. (12). In this way one obtains again a unique set of population, i.e. the canonical populations $P_n = e^{-\beta E_n} / \sum_m e^{-\beta E_m}$, where the parameter $\beta$ is given as implicit solution of the constraint for the expectation energy, $E = \sum_n E_n e^{-\beta E_n} / \sum_m e^{-\beta E_m}$.

We do not follow any of these procedures because it seems to us that there are no strong reasons to assign a superiority of a particular set of populations with respect to the other possible sets, neither to assign to them a particular functional dependence on the corresponding energy eigenstates or on some others parameters of the problem.

In our approach the equilibrium properties, defined according to the asymptotic time average, eq.(4), depend on the specific population set of the corresponding PSD, which in general is not the microcanonical ones and neither the canonical set. Thus, by excluding the exceptional case of a system prepared with a priori defined populations, it should be clear that a complete information on the populations is not available, and that one can analyze only statistically the different possibilities for the populations of the given system. In the next section, we shall consider ensembles of pure states, corresponding to different sets of populations, in order to specify a suitable probability distribution for them. This leads to the central problem considered in this paper which is the relation between possible population distributions and the thermodynamic characterization of the equilibrium properties of a single quantum system.

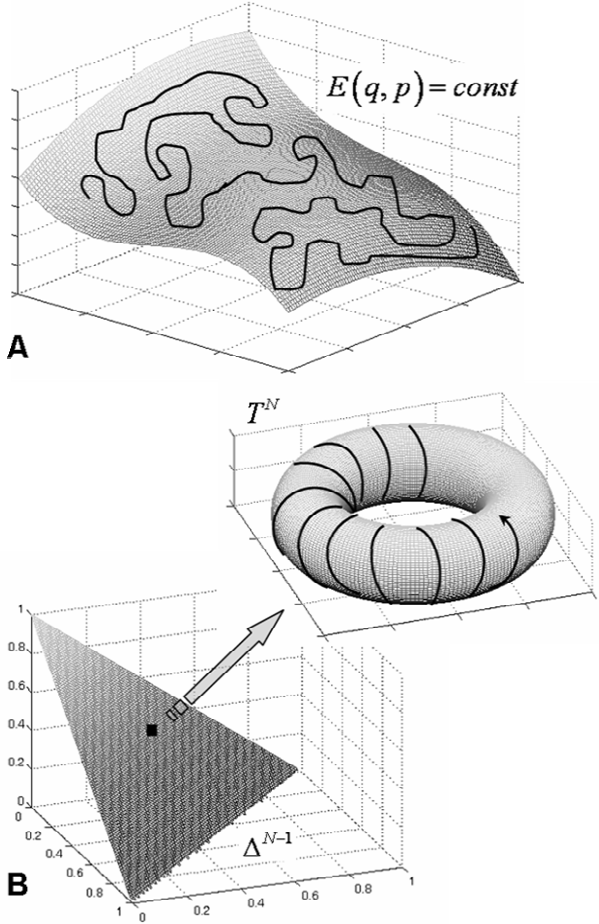

**Figure 1: Pictorial representation of the different dynamical evolutions in the classical (A) and in the quantum mechanical (B) framework. In the first case the ergodic hypothesis assume the constant energy surface to be densely filled by almost all the trajectories leading to the microcanonical distribution. On the contrary quantum dynamics is confined on the N-dimensional torus $T^N$ associated to the fixed population set belonging to the (N-1)-dimensional simplex $\Delta^{N-1} = \left\{ (P_1,...P_N) \in \mathbb{R}^N \left| \sum_{k=1}^N P_k = 1 \vee P_k \geq 0 \right. \right\}$, represented as a plane in the figure.**

## III. STATISTICAL ENSEMBLES

As the notion of *ensemble* dates back to the origins of statistical mechanics and can assume slightly different meanings, it is worth to specify how this concept is intended in the following. This is a very important point since different approaches to the justification of the distributions used in statistical mechanics largely depends on the different meanings attributed to the notion of statistical ensemble introduced in the original work of Gibbs [47,48,49].

The main object of our study is a single isolated quantum system described by its time dependent wavefunction with a particular set of populations. Thus no real ensemble of quantum systems is ever invoked. The concept of ensemble is used here to substantiate the notion of probability density for the populations specifying the pure state. On a phenomenological level, the ensemble could be identified with a statistical sample of the population sets (and, correspondingly, of isolated quantum systems). On a more formal ground [50], an ensemble shall be specified by the sample space $D$ for the possible population sets $P$ and the corresponding probability density (or ensemble distribution) $p(P)$, where the components of $P$ are the independent populations treated as stochastic variables.

Moving the focus from the single pure state to the statistical distribution of pure states introduces basically two problems. Firstly, i) one has to properly define the ensemble and different choices are possible, for instance in relation to the type of constraints which defines the sample space of the allowed populations. Secondly, ii) one has to establish the connection between the population distribution and the equilibrium properties observed in a given isolated system which have a uniquely defined set of populations.

With reference to the first point we shall consider two different ensembles. As mentioned in the introduction, similar ensembles have been used in previous works on typicality [3,12-20] in the form of a probability distribution on the entire wavefunction. On the contrary, according to our procedure, the ensembles specify the probability distribution of populations only, and they are associated to two conventional rules for selecting the sample space for the populations of the pure state:

1. *Random Pure State Ensemble* (RPSE): it includes all the normalized wavefunctions in a certain Hilbert sub-space, $\mathcal{H}_{RPSE} \subseteq \mathcal{H}$, which define the RPSE active space. Such an ensemble of random pure states is of special interest in the field of quantum information theory and their characteristic properties have been subject of several studies [51]. We select the RPSE active space according to a high energy cut-off $E_{max}$ for the allowed energy eigenstates

$$\mathcal{H}_{RPSE} = \text{span}\{|e_n\rangle | E_n \leq E_{max}\} \tag{25}$$

If one deals with an Hamiltonian with a bounded discrete spectrum, like in system of spins, one could consider the particular case of the whole Hilbert space, $\mathcal{H}_{RPSE} = \mathcal{H}$. If we would have introduced also a minimum energy condition on the basis of a parameter $E_{\min}$, by considering the Hilbert subspace spanned by the Hamiltonian eigenvalues within the energy shell $E_{\min} < E_n \leq E_{\max}$, then we recover the ensemble of wavefunctions considered by Goldstein et al. [12]. It should be emphasized that this is not equivalent to the quantum microcanonical statistics, as it is usually intended, as long as in the latter case the populations are not randomly distributed but, on the contrary, they have a univocally defined values.

2. *Fixed Expectation Energy Ensemble (FEEE):* all the $N$-dimensional normalized wavefunctions in the Hilbert space with the same value of the expectation energy are considered. This is analogous to the generalized microcanonical ensemble proposed by Brody at al. [10,21,22]. The interest on this ensemble basically arises from the observation that the expectation value of the Hamiltonian plays the same role in the Schrödinger equation, eq. (22), as the classical energy does in the Hamilton equation. Thus such an ensemble could be intended as the quantum counterpart of the classical microcanonical distribution.

These ensembles have been studied in Ref. [52], whose results relevant for the present work will be briefly summarized in the following. In order to find the probability distribution of the populations for these ensembles of pure states, it is useful to consider the inherent geometry of the Hilbert space. Indeed it is quite natural to assume that the probability of a certain set of wavefunctions is proportional to the measure of the set of their representative points in that space. Let us consider an element (wavefunction) $\psi$ of an $N$-dimensional Hilbert space and specify it through the vector $c \equiv (c^1, c^2, ..., c^N) \in \mathbb{C}^N$ of its components along the eigen-energy basis like in eq. (1). This complex vector can be written as a $2N$-dimensional real vector $x = (x^1, x^2, \cdots, x^{2N}) \equiv (\operatorname{Re} c, \operatorname{Im} c)$ in the real space $\mathbb{R}^{2N}$. Notice that in this representation the norm of the vector $\psi$ is given as the Euclidean norm

$$\|\psi\| = \sqrt{\langle \psi | \psi \rangle} = \sqrt{\sum_{j}^{2N} (x^j)^2} \tag{26}$$

and this identifies the metric tensor of the $\mathbb{R}^{2N}$ space as the unit tensor, $g_{ij}(x) = \delta_{ij}$. This also implies that the measure of any region is obtained by integration of the elementary volume element $dV = dx = dx^1 dx^2 ... dx^{2N}$. As described in detail in Ref. [52], by performing the change of variables $x \to (P, \alpha)$ one easily derives the metric tensor in the population-phase representation. Once the metrics is known, it is easy to specify the proper volume element of any surface embedded in such a space. Generally speaking, $n$ constraints on the coordinates of a $2N$ dimensional (real) space define an hyper surface of dimension $d = 2N - n$. In particular, the conditions of normalization and fixed expectation energy involve only the population coordinates and thus they define hypersurfaces which can be described by a set of local coordinates given by $P=(P_1,...P_{N-n})$ and $\alpha=(\alpha_1,...,\alpha_N)$, where $n=1$ if only the normalization is required (RPSE), while $n=2$ when considering wavefunctions with a common expectation energy (FEEE). Then, the proper volume element is specified as $dV=\sqrt{|g(P,\alpha)|}dPd\alpha$, where $g(P,\alpha)$ is the determinant of the metric tensor induced on the surface, $dP=dP_1...dP_{N-n}$ and $d\alpha=d\alpha_1...d\alpha_N$. Finally, from the assumed proportionality between the probability $p(P,\alpha)dPd\alpha$ and the measure $dV=\sqrt{|g(P,\alpha)|}dPd\alpha$, it follows that the ensemble probability density on the population is given as

$$p(P)=\frac{1}{\int_D dP\int d\alpha\sqrt{|g(P,\alpha)|}}\int d\alpha\sqrt{|g(P,\alpha)|} \qquad (27)$$

with normalization $\int_D p(P)dP=1$, where $D$ is the domain for the populations.

As long as the constraints which define an ensemble involve only populations, it turns out that the volume element does not depend on the phases, that is $\sqrt{|g(P,\alpha)|}dPd\alpha=\sqrt{|g(P)|}dPd\alpha$. This implies a uniform distribution of the phase variables [52] in agreement with the PSD, eq.(20), previously derived on the basis of the pure state evolution. By using the procedure described above, one finds [52] that for the RPSE the population sets are uniformly distributed on the $(N-1)$ dimensional simplex defined by the normalization condition eq.(11). The corresponding probability density on the $(N-1)$ independent populations reads

$$p_{RPSE}(P_1,...,P_{N-1})=(N-1)! \qquad (28)$$

Notice that such a distribution does not imply that the populations are statistically independent, because of the correlation between their domains of existence, due to the normalization eq. (11).

By including also the expectation energy, eq. (12), amongst the constraints one finds the following probability density on the $(N-2)$ independent populations of the FEEE

$$p_{FEEE}(P_1,...,P_{N-2})=\frac{1}{C}\left[\sum_{j=1}^{N-2}P_j(1+a_j)a_j-\left(\frac{E-E_{N-1}}{E_N-E_{N-1}}\right)^2+\left(\frac{E-E_{N-1}}{E_N-E_{N-1}}\right)\right]^{1/2} \qquad (29)$$

with coefficients $a_j=\frac{E_{N-1}-E_j}{E_N-E_{N-1}}$, and the constant $C$ determined by the normalization of the probability density.

In Ref. [52], it has been shown also that the joint distributions of populations, eq. (28) and eq. (29), is well approximated by a factorized probability distribution

$$p(P) \simeq \prod_{k=1}^{N} w^{(k)}(P_k) \tag{30}$$

where now the populations are considered as stochastic variables defined in the entire real positive axis. In eq. (30) the explicit forms of the single variable distributions $w^{(k)}(P_k)$ depend on the particular ensemble. For the RPSE one has

$$w^{(k)}(P_k) = N e^{-N P_k} \tag{31}$$

while for the FEEE

$$w^{(k)}(P_k) = \frac{(N-1)E_k}{E} e^{-\frac{(N-1)E_k}{E} P_k} \qquad k \neq 1$$

$$w^{(1)}(P_1) = \delta(P_1 - \langle P_1 \rangle) \tag{32}$$

where $\langle P_1 \rangle = 1 - \frac{E}{N-1} \sum_{k \neq 1}^{N} \frac{1}{E_k}$.

In the next section we shall face the second problem listed at the beginning of this section, that is, the relation between the population distribution and the equilibrium properties of a single quantum system. The following question has to be addressed: *Do they exist functions whose PSD average is approximately the same regardless of the specific set of populations considered?* Stated differently: *Are some functions of the populations, above all the thermodynamic functions, almost independent on the set of populations, but retaining a well defined dependence on the total energy of the system?* This property has been proposed as a good definition of "quantum ergodicity" [53], however this name might appear misleading as long as it has no relation with ergodicity as intended in classical mechanics. Thus in the following we shall avoid any reference to ergodicity in favors of the notion of the typicality of the thermodynamic functions in ensembles of pure states.

### IV. The Emergence of Typical Thermodynamic Properties

The problem of the connection between the quantum mechanical description and the characterization of the thermal equilibrium has been discussed in the past by many authors from different points of view [54,55]. One particular line of investigation emphasizes the role of "quantum chaos" in order to explain the emergence of thermal behavior in quantum systems [53,56,57,58]. More recently, the concept of typicality as the key to the emergence of standard statistical equilibrium behaviour has been discussed in various works [3,12,13]. In our framework the emergence of a typical value of an equilibrium property of an isolated quantum system can be established by the study of its ensemble distribution. In Ref. [52] it has been shown that for both the considered ensembles, the distribution of the entropy, eq. (13), approaches the Gaussian distribution with a variance decreasing as the dimension of the Hilbert space increases. This is readily explained as a manifestation of the Central Limit Theorem if one considers the entropy eq.

(13) as a stochastic variable which results from the sum of nearly independent terms, each of them depending on one population. This can be true for other functions which like the entropy have the structure of a "sum-function" with respect to the population variables.

The concept of typicality is employed with somehow different meanings in various fields like information theory [59], quantum mechanics [60], chemical [23] and statistical physics [3,12,13]. In order to clarify our use of such a concept, we recall its definition according to Volchan [61]. A generic form of typicality is associated to an event $A$ whose probability $\text{Prob}(A)$ is very close to unity: $\text{Prob}(A) = 1-\varepsilon$ for a very small $\varepsilon$. In other words the exceptions are very unlikely since the probability of the complementary event $\overline{A}$ is very small, $\text{Prob}(\overline{A}) = \varepsilon \ll 1$. In our framework we can consider a property $f(P)$ depending on the set $P$ of populations. For instance, this could be the entropy, the equilibrium value $\overline{a(t)}$, eq. (16), of an observable of the quantum pure state or even the amplitude of its fluctuations eq.(17). Once the ensemble has been chosen and the corresponding population distribution is derived, one can in principle compute the ensemble average $\langle f \rangle$ of the property $f$ as well as its variance $\sigma_f = \sqrt{\langle f^2 \rangle - \langle f \rangle^2}$. Then typicality will be assigned to property $f$ if

$$\sigma_f \ll \Delta f \tag{33}$$

where $\Delta f$ is the range for the possible values of the property. Thus, the typical event is assigned to $f$ falling in an interval centered on $\langle f \rangle$, like $\langle f \rangle - 2\sigma_f \leq f \leq \langle f \rangle + 2\sigma_f$, much smaller than the full range of its possible values. In this case the average $\langle f \rangle$ will be called the typical value of the property $f$. In this sense typicality can be considered as the conceptual bridge between the behaviour of the single quantum system and the ensemble point of view of the statistical mechanics, in the meaning that for many properties of interest, it does not matter our impossibility to know the state of the system in detail, just for the remarkable fact that almost all quantum states behave essentially in the same way. Indeed, if condition (33) holds, the ensemble average $\langle f \rangle$ becomes a good estimate of the property $f$ for the vast majority of pure states of the ensemble. In part II of the present work, we shall employ the condition eq. (33) to verify the typicality of the functions of interest in the considered model system.

Definition of typicality according to eq. (33) is somehow vague for the lack of a quantitative criterion, as long as the probability of the exceptional set $\text{Prob}(\overline{A}) = \varepsilon$ is left to a subjective choice. A stronger form of typicality can be assigned to an almost sure event $A$ with a unitary probability

$$\text{Prob}(A) = 1 \tag{34}$$

An example is the probability of getting an irrational number from a random choice within the $[0,1]$ interval of real numbers, since the measure of the set of the exceptions is null. In the present framework the almost sure typicality, according to eq. (34), is assign to a property when the ratio $\sigma_f / \Delta f$ tends to zero in the limit of an increasing dimension of the Hilbert space $N$

$$\lim_{N \to \infty} \frac{\sigma_f}{\Delta f} = 0 \qquad (35)$$

in relation to the thermodynamic limit of the system. In such a limit the probability that $f$ falls within a finite fraction of its domain centered in $\langle f \rangle$ tends to unity.

It is evident that if the variance $\sigma_f$ can be evaluated as a function of the Hilbert space dimension (or the number of components of the system), both the forms of typicality, i.e. the weaker form of eq. (33) and the typicality of the almost sure event eq. (34), can be employed. The latter is useful to establish the thermodynamic limit of a given type of systems, while the former has to be applied in the analysis of finite systems, in particular for establishing whether the ensemble average $\langle f \rangle$ is representative of a statistical sample of a given property.

The existence of a typical value for a large class of observables can be viewed as a manifestation of the "concentration of measure" phenomenon [62]: it is a striking fact from elementary geometry of high dimensional surfaces that the uniform measure on the $k$ dimensional sphere, $S^k$, is strongly concentrated about any equator when $k$ gets large; consequently any polar cap strictly smaller than a hemisphere has a relative volume exponentially small in $k$. This induces a similar behavior for any slowly varying function on the sphere, which we can understand indeed as a random variable induced by the uniform measure on the sphere: namely, it will take values close to the average except for a set of volume exponentially small in $k$. This idea becomes rigorous in the Levy's Lemma which is also the main ingredient in the general proof of the typicality given in Popescu et al. [3]. Since pure quantum states which lies in a Hilbert space of dimension $N$ can be represented as real unit vectors in a $2N$-dimensional phase space, the above observations on the sphere ensure that, as the dimension of a quantum system becomes large, the behaviour of the typical value of a certain property of the quantum state becomes meaningful.

Besides the entropy, also the internal energy $U$ has to be assigned in order to recover a thermodynamical description of a quantum system. An obvious choice would be the identification of the internal energy with the expectation value $E$ of the quantum Hamiltonian. Such a simple choice, $U = E$, can be adopted for the FEEE having the expectation energy as an independent parameter. Thus, once the typicality of the entropy has been verified, one can derive the (thermodynamic) state function displaying the internal energy dependence of the entropy, by calculating the average entropy $\langle S \rangle$ as a function of the expectation energy $E = U$, i.e. the function $\langle S \rangle (U)$. A more complex scenario has to be invoked for the RPSE, since in this case the expectation energy is not an independent parameter and depends on the population set describing a given pure state. Therefore, such an ensemble supplies a distribution of expectation energies $E$ and, in order to recover a meaningful correspondence between internal energy and expectation energy, we have to assume typicality also for such a property. If this condition is verified, then the internal energy can be identified with the ensemble average $\langle E \rangle$ of the expectation energy:

$$U = \langle E \rangle \tag{36}$$

It should be recalled that RPSE has the high energy cutoff, $E_{max}$, as the only independent parameter. Then the ensemble averages of the entropy and of the expectation energy have both to be considered as functions of such a parameter, that is $\langle S \rangle (E_{max})$ and $U(E_{max})$. By eliminating the $E_{max}$ dependence from these two functions, one obtains the (thermodynamic) state function $\langle S \rangle (U)$ for the internal energy dependence of the entropy also for the RPSE.

Given the state function $\langle S \rangle (U)$ describing the dependence of the entropy on the internal energy, the other thermodynamic parameters are easily recovered. In particular we focus on the absolute temperature which is derived through the derivative of such a function

$$\frac{1}{T} = \frac{d \langle S \rangle (U)}{dU} \tag{37}$$

However, the existence of a state function $\langle S \rangle (U)$ for the entropy is not sufficient alone to assure the congruence between the statistical properties of an ensemble and the macroscopic properties described by the thermodynamics. Further requirements have to be verified, in particular that $\langle S \rangle (U)$ is a convex increasing function of $U$. Only in this case one recovers from eq. (37) a temperature increasing with the internal energy. Furthermore both $\langle S \rangle$ and $U$ must behave like extensive parameters, in order to recover from eq. (37) a temperature with an intensive character.

Besides the congruence with macroscopic thermodynamics, it is important also to assure that the analysis of ensembles allows one to recover the canonical statistics for a system in contact with a thermal bath. Thus we have to verify that the elements of the equilibrium reduced density matrix $\bar{\mu}$ of a subsystem, eq. (15), become typical as the system size increases. If such a condition holds, then the canonical statistics would be recovered if the ensemble average of the equilibrium density operator of the subsystem can be written as

$$\langle \bar{\mu} \rangle = \exp(-H_S / k_B T) / \text{Tr}_S \{\exp(-H_S / k_B T)\} \tag{38}$$

where $H_S$ denotes the subsystem Hamiltonian.

It should be emphasized that the temperature defined by the thermodynamic relation eq. (37) for the overall system, which we shall call the global temperature, and the temperature which appears in the canonical equilibrium state of a subsystem eq. (38), which we shall term the local temperature, must be the same on a physical ground. However, such a result should not taken for granted, in the meaning that it has to be considered as a requirement for the correct definition of the statistical ensemble of populations.

In conclusion five major requirements have to be verified in order to recover from statistical ensembles a physically reasonable representation of real systems:

1) Typicality of $S$ and $E$.
2) $\langle S \rangle = \langle S \rangle (U)$ as a convex increasing function of $U \equiv \langle E \rangle$.
3) Both $\langle S \rangle$ and $U$ should be extensive for systems large enough.
4) Typicality of the time-averaged density operator $\bar{\mu}$ of a subsystem in correspondence of its canonical form.
5) Equivalence between the global and the local temperatures.

In part II we will investigate the relevant properties of a spin model system in order to verify whether the requirements 1-5 are fulfilled in the previously introduced ensembles (i.e., the FEEE and the RPSE).

## V. Summary and Conclusion

Having identified the equilibrium properties of a quantum pure state, eq. (20), as the main objective of our study, we have introduced the Pure State Distribution, eq. (20), in the space described by the coordinates $(P, \alpha)$. This distribution reflects the features of the temporal evolution ruled by the Schrödinger equation: in the $2N$ dimensional phase space one has $N$ constants of the motion, i.e. the populations, while for time long enough the phase variables take all their possible values with a uniform probability distribution as long as the eigenenergies are rationally independent.

To conciliate the idea of an equilibrium state defined on the basis of the dynamics of the system with the fundamental notion of thermal equilibrium, one should study the distributions of the thermodynamic functions in statistical ensembles of pure states. Our setup allows the analysis of the role of typicality in the emergence of the equilibrium statistical thermodynamics within a simple but effective theoretical framework.

Recent approaches [3,12,13] to the problem of relating the statistical ensemble to the property of a single pure state on the basis of the concept of typicality have been of pure "geometrical" character, by analyzing the distribution of the possible wavefunctions describing an isolated system, without taking explicitly into account the role of the evolution of the quantum state. In our analysis both the ingredients are present, that is the quantum dynamics which determines the equilibrium properties on the one hand, and the geometrical statistics for the populations on the other hand. However they play different roles in determining the statistics of quantum states and this is emphasized by the introduction of the two different probability distributions: the Pure State Distribution on the one hand and the population distribution characterizing the ensemble of pure states on the other hand.

Another major difference with respect to previous typicality based approaches [3,12,13] is that we consider explicitly not only subsystem properties described by the reduced density matrix, but also thermodynamic properties of the overall isolated system, as the energy and the entropy. Even these quantities depend on the pure state realization, that is on the population set. Therefore the issue of their typicality has to be risen in order to justify the thermodynamic description of an isolated quantum system.

Our analysis does not assume from the beginning a uniquely defined statistical ensemble for the populations. On the contrary we think that a priori rules for the statistics of pure states are not available, besides those imposed by quantum dynamics which leads to the Pure State Distribution. Therefore different statistical ensembles for populations can be proposed, and the selection amongst them has to be done on the basis of the agreement with macroscopic thermodynamics and the canonical behaviour of subsystems. In the second part of this work, such an analysis will be presented for the two previously introduced ensembles (RPSE and FEEE) in relation to a specific spin model. Afterwards we will draw the general conclusions of the overall investigation.

**Acknowledgements**

The authors acknowledge the support by Univesità degli Studi di Padova through 60% grants.

**Appendix A: Derivation of the PSD**

In order to prove that the homogeneous distribution on the phases eq. (20) is the distribution which satisfy the equivalence eq. (19) with the asymptotic time average, let us introduce the Fourier expansion of the distribution function

$$p(P|\alpha) = \sum_{n_1} \exp(in_1\alpha_1) \sum_{n_2} \exp(in_2\alpha_2) \cdots \sum_{n_N} \exp(in_N\alpha_N) p_{n_1,n_2,\ldots n_N}(P) \tag{A1}$$

each coefficient being given as

$$p_{n_1,n_2,\ldots n_N}(P) = \frac{1}{(2\pi)^N} \int d\alpha \exp\left(-i\sum_{j=1}^{N} n_j \alpha_j\right) p(P|\alpha) \tag{A2}$$

that is, the phase average

$$p_{n_1,n_2,\ldots n_N}(P) = \overline{f} \tag{A3}$$

of the function

$$f(\alpha) = \frac{1}{(2\pi)^N} \exp\left(-i\sum_{j=1}^{N} n_j \alpha_j\right) \tag{A4}$$

On the other hand, by evaluating $\bar{f}$ as the average along the trajectory, and by using eq. (9) for the time dependence of the phases, the following relation is recovered

$$\bar{f} = \lim_{T\to\infty} \frac{1}{T}\int_0^T dt\, f(\alpha(t)) = \lim_{T\to\infty} \frac{1}{T}\int_0^T dt\, \frac{1}{(2\pi)^N} \exp\left(-i\sum_{j=1}^{N} n_j \alpha_j(t)\right) =$$

$$= \frac{1}{(2\pi)^N} \exp\left(-i\sum_{j=1}^{N} n_j \alpha_j(0)\right) \lim_{T\to\infty} \frac{1}{T}\int_0^T dt\, \exp\left(-\frac{i}{\hbar} t \sum_{j=1}^{N} n_j E_j\right) \tag{A5}$$

This quantity is not vanishing only if the following constraint is satisfied

$$\sum_{j=1}^{N} n_j E_j = 0 \tag{A6}$$

Then, by comparing eq. (A3) with eq. (A5), we derive that an expansion coefficient $p_{n_1,n_2,...n}(P)$ is not vanishing only under the previous condition eq. (A6). An obvious case is that for vanishing values of all the indices, $n_1 = n_2 ... = n_N = 0$. If only this trivial solution exists the values of the eigenenergies are said to be rationally independent. It is well known that under such a condition the trajectories of the phase variables densely fill the N-dimensional torus [29]. Thus, by excluding the situation of rationally dependent energy levels, the constraint eq. (A6) implies that all the indices are vanishing. Therefore, the expansion coefficients are given as

$$p_{n_1,n_2,...n_N}(P) = \frac{\prod_{j=1}^{N} \delta_{n_j,0}}{(2\pi)^N} \tag{A7}$$

and this result, according to the Fourier expansion eq. (A1), corresponds to the homogeneous distribution eq. (20). It should be noted that the previous analysis requires that none of the eigenvalues is vanishing. If, for instance, the first eigenvalue is null, $E_1 = 0$, then condition eq. (A6) is realized whatever is the integer number $n_1$. On the other hand, the scale of energy can always be shifted, so to force the vanishing of one eigenvalue. In such a case the corresponding phase is independent of time according to eq. (9), $\alpha_1(t) = \alpha_1(0) = \alpha_1^0$, that is $\alpha_1$ becomes a constant of motion. For the other phases one can still apply the previous analysis based on their independence. The resulting probability density on the phases

$$p(P|\alpha) = \delta(\alpha_1 - \alpha_1^0) \frac{1}{(2\pi)^{N-1}} \tag{A8}$$

albeit different from that of eq. (20), leads to the same values for the average of observables eq. (3). Indeed, since the elements of the density matrix eq. (10) depend on the difference between two phases, by performing the change of variables $\alpha_n \to \alpha_n' = \alpha_n - \alpha_1$ for $n = 2, 3, \cdots, N$, the density matrix becomes independent of the first phase $\alpha_1$ and, therefore, the distributions eq. (20) and eq. (A8) produce the same averages.

## APPENDIX B: The Birkhoff's theorem

The main objective of the statistical mechanics of classical isolated systems is the replacement of the mechanical description given by a trajectory, with a description in terms of probability density on the space which represents the possible states of the evolving system during its motion [63], i.e. the phase space $\Gamma$. The ergodic approach to the foundation of statistical mechanics allows one to recognize the conditions of validity of such a replacement, which we briefly recall in the following:

1. There are subspaces of the phase space which always transforms into themselves during the natural motion, and they are called *invariant parts* of the phase space.
2. An invariant part $V$ is called *metrically indecomposable* if it cannot be represented in the form $V = V_1 + V_2$ where $V_1$ and $V_2$ are in turn invariant parts with non vanishing measure.
3. Birkhoff's Theorem states that, by letting $V$ be an invariant part of $\Gamma$ with finite volume and $f(Q)$ a phase function defined at all points $Q \in V$, and given the trajectory $Q(t)$ of a point in the phase space, then the following limit

$$\bar{f} = \lim_{T \to \infty} \frac{1}{T} \int_0^T f(Q(t)) dt \tag{B1}$$

exists almost everywhere on $V$. Moreover if $V$ is metrically indecomposable, then almost everywhere on $V$

$$\bar{f} = \frac{1}{\mathcal{M}(V)} \int_V f(Q) dV \tag{B2}$$

where $\mathcal{M}(V)$ is the total measure of the set $V$

$$\mathcal{M}(V) := \int_V dV \tag{B3}$$

Then $\bar{f}$ can be identified with the statistical average $\bar{f} = \int_V f(Q) p(Q) dV$ with a constant probability density given as $p(Q) = 1/\mathcal{M}(V)$.

The proofs of these results as well as a detailed discussion of their implication in classical statistical mechanics can be found in Khinchin's book [63]. The Birkhoff's theorem is important for the foundations of statistical mechanics because it provides a rigorous proof of the equivalence between asymptotic time averages, which represent *by definition* the equilibrium properties of the system, and phase space averages. The equivalence can be established if the time evolution of the system, i.e. the motion of its representative point in the phase space, covers all the region of phase space where we want to perform the averaging procedure according to eq. (B2).